\documentclass[12pt,eqsecnum,preprint]{aastex}
\begin{document}
\title{Force Free Waves and Black Hole Magnetospheric Causality}
\author{Brian Punsly}
\affil{4014 Emerald Street No.116, Torrance CA, USA 90503}
\email{brian.punsly@hsc.com or brian.punsly@gte.net}
\begin{abstract}The force free approximation is often useful when
  describing tenuous plasmas in strong cosmic magnetic fields. Time
  evolution of any such system is governed by the information that can
  be transported along the characteristics of the plasma modes allowed
  by the force free constraint. This brief article elucidates, for the
  first time, the nature of the information that can be transported along the
characteristics for each of the two plasma modes, the Alfven and fast
  modes. It is reassuring that these modes behave as one would expect
  if the perfect magnetohydrodynamic (MHD) modes in a magnetically
  dominated plasma were evaluated in the limit of zero plasma inertia
  (i.e., only allowed to propagate information consistent with the
  force free constraint). There are no properties of these waves that
  do not exist within the MHD theory, when this additional constraint on the current is imposed. The results of the
  characteristic analysis elucidates the nature of the causality
  violation in force free black hole magnetospheres. The most
  significant result from the standpoint of global causality is that
  charge and current perturbations can not be transported along the
  fast mode characteristics in the force free limit. 
\end{abstract}
\section{Introduction}
Applications of force free electrodynamics are
  common in  the astrophysical literature. The assumption that the
  plasma inertia is negligible compared to the magnetic energy density
  is an accurate depiction of the plasma state in many contexts. This
  condition justifies the force free approximation based on a local
  physical frame analysis. The most notable successes of such an
  approach are descriptions of equilibria in the solar
  magnetosphere. Motivated by the solar work, it has also been
  extended to studies of the physics of black hole magnetospheres (see
  \citet{blz77} and \citet{thp86}) and more recently gamma ray bursts (see \citet{bla02}).
\par However, the implementation of the force free condition to black
  hole magnetospheres is not so trivial because general relativity is
  not a local theory. In particular, the concepts of an event horizon
  and ergosphere are globally defined \citep{haw73}. Black hole
  magnetospheres are the main focus of this treatment, \textbf{yet all of the
  results that are derived in this article apply equally well to any force free calculation}. A salient example of
  the global nature of general relativity is the fact the physical
  boundary condition at the event horizon is that all particles that
  are near the black hole must corotate with the event horizon and go
  into the black hole on trajectories that approach the ingoing light-like
  principal null congruence, as seen by any external observer (see
  Chapter 3 of \citet{pun01}). This applies to any plasma, no matter
  how tenuous and is independent of the magnetic field strength. There
  are no stable equilibria near the horizon, gravity always overwhelms
  the other forces. This property is precisely what distinguishes a
  black hole from another type type of compact object such as a
  neutron star or white dwarf. All plasma near the horizon is
  inertially dominated: a global condition. In a sense, the plasma
  inertia is infinite near the horizon as viewed globally. By
  contrast, a local analysis would lead one to believe that the force
  free approximation is valid (i.e., the plasma inertia is
  negligible). In spite of  this defining property of the black hole,
  force free treatments of black hole magnetospheres that extend all
  the way to the event horizon have been proposed by many authors (most notably \citet{blz77} and \citet{thp86}).
\par In this paper, we consider the plasma waves in a force free
  magnetosphere. This is important in order to clarify certain
 anecdotal defenses of the force free approximation near the event
 horizon which infer that force free plasma waves can transport
 different information than their counterparts in the full perfect
 magnetohydrodynamic (MHD) treatment of plasma waves in magnetically
 dominated magnetospheres  (see Chapter 2 of \citet{pun01} for a detailed MHD
 treatment and \citet{bla01} for the force free discussion). In this
 article, it is shown by means of a straightforward explicit
 calculation that this claim is erroneous. This article elucidates the
 current structure within the individual wave modes for the first time
 (even the lengthy, complicated and very detailed analysis of certain
 topics of \citet{uch97} in the context of black hole magnetospheres
 does not address these most relevant issues in a force free magnetosphere).
\section{The Force Free Approximation and Magnetically Dominated
 MHD}This section compares and contrasts the force free approximation
with a magnetically dominated perfect MHD system. Generally, the force
free approximation is considered as a zero mass limit of the full MHD
theory. It is usually considered an expedience that reduces the dimension
of the full MHD system of hyperbolic differential equations to
five. It is very rarely argued to be a more accurate physical
description than the full MHD theory (but, see \citet{bla02} for such
an argument). Generally, the force free approximation is invoked is to
simplify the mathematical treatment not to improve on the physical
depiction. If the force free approximation is not pathological then one
would expect it to occur as the natural limit of zero inertia in the full
MHD theory. This section begins to lay the groundwork to show that
this is indeed the case. The set of force free equations are the
current constraint written covariantly in terms of the Maxwell tensor
and in a local inertial coordinate system,
\begin{mathletters}\begin{eqnarray}
            && F^{\mu\nu}J_{\nu}=0\;,\\
&& \rho_{e}\mathbf{E} +\frac{\mathbf{J}\times \mathbf{B}}{c} = 0\;,\end{eqnarray}\end{mathletters}
and Maxwell's equations (written in local coordinates) 
\begin{eqnarray}
 &&\mathbf{\nabla}\times\mathbf{B}=\frac{1}{c}
\frac{\partial \mathbf{E}}{\partial t}+\frac{4\pi}{c}\mathbf{J}\;,\\
&& \mathbf{\nabla}\times\mathbf{E}= -\frac{1}{c}
\frac{\partial\mathbf{B}}{\partial t} \;,\\
&& \mathbf{\nabla}\cdot\mathbf{B}=0 \;,\\
&& \mathbf{\nabla}\cdot\mathbf{E}=4\pi \rho_{e} \; .
\end{eqnarray}
Equation (2.1) also implies the conditions:
\begin{eqnarray}
&& \mathbf{E}\cdot\mathbf{B}=0 \;, \\
&& \mathbf{E}\cdot\mathbf{J}=0\;.
\end{eqnarray}
The force free condition in equation (2.1) arises from the zero mass
density limit of perfect MHD, the plasma has no inertia to stress the magnetic field lines.
\section {Force Free Plasma Waves}The causal structure of a force free
system is governed by the waves supported in the force free plasma. In
this section we derive the nature of the information that can be
transported along each of the wave characteristics. Recall that in perfect MHD the Alfven mode can transport charge and current that was either field aligned
(force free) or cross-field (inertial). The fast wave cannot transport
charge or field aligned currents, only inertial currents (see
\citet{pun01}, Chapter 2). It turns out that the effects of the force
free constraint are equivalent to the zero mass density limit of
perfect MHD plasma waves, as expected. Since the fast wave
still cannot transport force free current in the zero inertia limit, when
the force free constraint on the current is imposed, the only value of
current consistent with these two concepts (effectively two
mathematical equations) is zero current. The fast mode in the force free limit carries no current or charge.
\par As a consequence of equation (2.6), one can find a frame in which the
electric field vanishes if the Lorentz invariant
$F^{\mu\nu}F_{\mu\nu}>0$ (i.e., the field is magnetic). There exists a
coordinate transformation to this frame that resembles a Lorentz boost in the
azimuthal direction in an axisymmetric magnetosphere except that the
boost parameter can exceed the speed of light. Thus, this is not a
Lorentz transformation in general. However, the angular velocity as
observed from asymptotic infinity in an asymptotically spherical
coordinate system is referred to as the field line angular velocity,
$\Omega_{_{F}}$, even when the azimuthal velocity is faster than the
speed of light \citep{bla02,pun01}. Thus, this frame is often called
the corotating frame of the magnetic field. The light cylinder is
defined as the surface at which the azimuthal boost velocity equals
the speed of light (in all physical frames). The implementation of
this frame facilitates the calculations and one can perform a general
coordinate transformation in order to evaluate plasma quantities in  other
physical frames. Since, this is not a Lorentz transformation in
general, it is best to find coordinate invariant expressions. Another
frame in which the electric field vanishes is the plasma rest frame. However,
this frame is poorly defined in the force free limit since there is no
accounting of the plasma state. Frames in which the electric field
vanishes are sometimes called proper frames.
\par The calculations are performed in the short wavelength
approximation. This is equivalent to the analysis found in
\citet{bla02}. No matter what the radius of curvature of the magnetic
field, one can find  plasma mode wavelengths that are short compared to
this value. This is not very restrictive, even close to the event horizon of
a black hole. One should be able to find proper frames in which the
radius of curvature of the magnetic field is on the order of the
geometrized mass of a black hole (~$10^{14}$ cm for a supermassive black
hole). The unperturbed value of the four current, $J^{\mu}_{0}$,
effectively vanishes in the short wavelength approximation that the
waves vary on much smaller scales than the background magnetic fields
(the currents are effectively created by the gradients of the magnetic
field). As a result of Ampere's law, this approximation is equivalent
to treating the magnetic field of the unperturbed state as approximately a constant in the proper frame,
\begin{eqnarray}
    &&\frac{\partial}{\partial x^{\alpha}} \mathbf{B}\approx 0 \;.
    \end{eqnarray}
\par We analyze short wavelength linear (small amplitude)
perturbations of equations (2.1)-(2.7). The waves are chosen in a plane wave representation to have an oscillatory behavior, $\exp i(\mathbf{k} \cdot
\mathbf{r}-\omega t)$. We define the $x$ direction to be parallel to
$\mathbf{k}$, so all perturbed quantities are Fourier analyzed in
$\left(x, t \right)$  space. For example, the four current is
\begin{mathletters}\begin{eqnarray}
&&
J^{\mu}= J_{_{0}}^{\mu}+e ^{i(kx-\omega t)}\delta \! J^{\mu} \\
&&
\quad \;\, \approx e^{i(kx-\omega t)}\delta \! J^{\mu} \; .
\end{eqnarray}\end{mathletters}
\par The interested reader can find a discussion of the effects of field
line curvature in a black hole magnetosphere in Chapter 6 of
\citet{pun01}. The nature of the resultant hybrid modes is beyond the
scope of this treatment that has been chosen to parallel the analysis
of \citet{bla02}. In summary, field line curvature mixes the properties
of the plasma modes as it does in the earth's magnetosphere \citep{lan79}. Long
wavelength modes are neither fast nor Alfven modes, but have hybrid
properties (Appendix A is a derivation of the first order effects of
magnetic field line curvature in the force free limit). The faster
mode propagates slower as the wavelength increases and is no longer related
to the critical point structure of the winds that reflect short
wavelength mode speeds. The implication for black hole magnetospheres
is that long wavelength force free fast modes travel slower than the speed
of light and therefore can not propagate information from near the
event hroizon to asymptotic infinity.
\par The analysis of long wavelength plasma modes is very complicated
and one is referred to the in depth analysis in Chapter 6 of \citet{pun01}
to see the details of outgoing mode propagation resulting from field
line curvature and spacetime curvature near the event horizon of a
black hole. The results of those calculations are in accord with the
causality arguments in this article. In particular, one finds that
long wavelength plasma modes are extremely inefficient in their
ability to transport magnetic stresses, field aligned current and
charge outward from the spacetime near an event horizon.
\par In a frame with a vanishing electric field, the perturbed
equations (2.1) - (2.7), with the condition above become:\begin{eqnarray}
&& \mathbf{B}\times\delta \mathbf{J}=0\;,\\
&& \omega\delta \mathbf{E}=-c\mathbf{k}\times\delta \mathbf{B}- i 4\pi\delta\mathbf{J}\;,\\
&& \omega\delta \mathbf{B}=c\mathbf{k}\times\delta \mathbf{E} \;,\\
&& \mathbf{k}\cdot \delta \mathbf{B}=0\;,\\
&& i\mathbf{k}\cdot \delta \mathbf{E}= 4 \pi \delta\!\rho_{e}\;,\\
&& \delta\mathbf{E}\cdot\mathbf{B}=0\;.
\end{eqnarray} It is useful to combine (2.2) and (2.3) to get the second order Maxwell's equation,
\begin{eqnarray}
    &&
    \mathbf{\nabla}^{2}\mathbf{E}-\mathbf{\nabla}(\mathbf{\nabla}\cdot
    \mathbf{E})=\frac{1}{c^{2}}\frac{\partial^{2}\mathbf{E}}
    {\partial t^{2}}+\frac{4\pi}{c^{2}}\frac{\partial
    \mathbf{J}}{\partial t} \; .
    \end{eqnarray}
This equation gives the following relation amongst perturbed electromagnetic quantities,
 \begin{eqnarray}
  &&
  k^{2}\delta \mathbf{E}- \mathbf{k} \left (\mathbf{k}\cdot \delta \mathbf{E}\right)
  =\frac{\omega^{2}}{c^{2}}\delta \mathbf{E}+\frac{i 4\pi\omega}{c^{2}}\delta\mathbf{J} \; .
    \end{eqnarray}
At this point we explicitly write out the components of the
 perturbed equation above in a local Cartesian basis. The x-axis was
 previously defined to be parallel to $\mathbf{k}$ and we
 define the z-axis such that $\mathbf{B}$ always lies in
 the x-z plane. The components of the second order Maxwell's equation become:
 \begin{mathletters}\begin{eqnarray}
&&{\delta \! E_{y}=\frac{4 \pi i \omega}{\left(c^{2}k^{2}-\omega^{2}
     \right)}\delta \! J_{y}\;,}\\
&&\delta \! E_{z}=\frac{4\pi i
  \omega}{\left(c^{2}k^{2}-\omega^{2}\right)}\;\delta \! J_{z} \; ,\\
&&\delta \! E_{x}=-\frac{4\pi i}{\omega}\;\delta \! J_{x} \; .
 \end{eqnarray}\end{mathletters}
\par Similarly an expansion of the components of the perturbed force
 free condition in (3.2) yields the following two conditions on the
 current:
\begin{eqnarray}
&& \delta \! J_{y}=0\;,\\
&& \delta \! J_{z} = \tan{\theta}\,\delta\! J_{x}\;,
\end{eqnarray} where $\theta$ is the angle between the propagation vector and the magnetic field direction.
There is also a third constraint on the current that is obtained by
substituting the perturbed second order Maxwell's equation (3.11) into the perturbed degeneracy condition (3.8):
\begin{eqnarray}
&& \delta \! J_{x} \left(c^2 k^2-\omega^2\right) =
\omega^2\frac{k_{\perp}}{k_{\parallel}}\delta\! J_{z}\;,
\end{eqnarray} where $k_{\parallel}$ and $k_{\perp}$ are the
components of the propagation vector parallel and perpendicular to the magnetic field direction, respectively.
\par Equation (3.14) can be substituted into the relation (3.13) in order to eliminate $\delta\! J_{x}$.
This yields the dispersion relation for plasma waves in a force free medium,
\begin{eqnarray}
&&\left[\omega^2 -c^2 k^2\right]\left[\omega^2 - c^2 k_{\parallel}^2\right]=0\;.\end{eqnarray}
\subsection{The Fast Mode}
The dispersion relation (3.15) has two solutions for the phase
velocity. The fast mode is characterized by the phase velocity,
\begin{eqnarray}
&& \frac{\omega}{k}=\pm c\;.
\end{eqnarray}
The group velocity is
\begin{eqnarray}
&& \frac{\partial\omega}{\partial\mathbf{k}}=\pm c\frac{\mathbf{k}}{k}\;.
\end{eqnarray}
Inserting the value of the phase velocity into the perturbed induction
equation, (3.5), and the perturbed degeneracy condition, (3.8), along with
the vanishing divergence of the magnetic field, (3.6), and the
constraint on the perturbed current, (3.14), we can solve for the
electromagnetic components of the fast wave:
\begin{mathletters}\begin{eqnarray}
&& \delta\! B_{x}=0\;,\\
&& \delta\! B_{y}=0\;,\\
&& \delta\! B_{z}=\delta\! E_{y}\;,\\
&& \delta\! E_{x}=0\;,\\
&& \delta\! E_{y}=\delta\! B_{z}\;,\\
&& \delta\! E_{z}=0\;.
\end{eqnarray}\end{mathletters}
Substituting the values of the electric field from (3.18d-f) into the second order Maxwell's
equation, (3.11), and the perturbed Gauss' law, (3.7), the light-like
phase velocity of the fast wave requires that \textbf{all of the components of
the four-current density that are propagated along the characteristics of the force free fast wave vanish identically},
\begin{eqnarray}
&& \delta\! J^{\mu}=0\;.
\end{eqnarray}This is a coordinate invariant expression and is
therefore true in all coordinate systems. The fast mode is linear in
the force free limit, since there is a unique wave propagation speed
given by (3.17). Thus, fast wave packets are linear superpositions of
the oscillatory solutions. Therefore, the four-current density
vanishes for fast wave packets as well.
\par From (3.18), one also has the following coordinate
independent statements: \begin{mathletters}\begin{eqnarray}
&& \ast\delta\! F^{\mu\nu}\delta\! F_{\mu\nu}=0\;,\\
&&\delta\! F^{\mu\nu}\delta\! F_{\mu\nu}=0\;,\\
&&\ast\delta\! F^{\mu\nu} F_{\mu\nu}=0\;,\\
&& \delta\! F_{\mu\nu}U^{\nu}_{_{G}}=0\;,\\
&& \frac{\omega}{c}U^{\nu}_{_{G}}=k^{\nu}\;,
\end{eqnarray}\end{mathletters}
where $U^{\nu}_{_{G}}$ is the group four velocity of the wave, which
is expressed in the coordinates of the proper frame as:
\begin{eqnarray}
&& U^{\nu}_{_{G}}\equiv \left( c,\;\frac{\partial
    \omega}{\partial\mathbf{k}}\right)\;.
\end{eqnarray} Not all of the relations in (3.20) are independent, for
    example (3.20d) implies (3.20a).
\par The primary result of this analysis, (3.19), that demonstrates
the vanishing of the current in a fast mode is not a result of a
biased interpretation by this author, a miscalculation or an unfounded
assumption. The same result actually follows from the results of
\citet{bla02}. First of all, the wave structure that is described in a
coordinate independent manner in (3.20) can be rewritten in the
``laboratory frame coordinates'' used by \citet{bla02}. One then finds
that, $\mathbf{k}$, $\delta\mathbf{B}$ and $\delta\mathbf{E}$ form an
orthogonal triad as well as equations (7) and (8) of \citet{bla02}. These are
the results that are used in that treatment to describe the causal
relevance of a fast wave to a force free magnetosphere. Inserting (7)
and (8) into Ampere's law, either (2.2) or (3.4), yields the fact that
$\delta J_{y}=0$ and $\delta J_{z}=0$. By the orthogonality of the
triad, $\delta E_{x}=0$, as a consequence of the definition of the propagation direction. Then Ampere's law can
be used again to show that $\delta J_{x}=0$. Finally, the vanishing of the
charge density follows from either Gauss' law or current conservation. Thus,
(3.19) can be found from the analysis of \citet{bla02} quite readily.
\subsection{The Alfven Wave}The other root of the dispersion relation
    (3.15) gives the phase velocity of the of the Alfven wave:
\begin{eqnarray}
&& \frac{\omega}{k} =\pm  c\, \cos{\theta}\;.
\end{eqnarray}One can also compute the group velocity,
\begin{eqnarray}
&& \frac{\partial\omega}{\partial\mathbf{k}}=\pm c \frac{\mathbf{B}}{B}\;.
\end{eqnarray}Thus, the Alfven mode also propagates information at the speed of light, but only along the
direction of the magnetic field.
\par Combining, the vanishing divergence of the perturbed magnetic
field, (3.6), along with the constraint on the perturbed current,
(3.14), and the induction equation, (3.5), inserted into the second
order Maxwell's equation, (3.11), one can solve for the components of
the perturbed electromagnetic field transported along the characteristics of the Alfven mode:
\begin{mathletters}\begin{eqnarray}
&& \delta\! B_{x}=0\;,\\
&& \delta\! B_{y}=- i\frac{4\pi k}{ck_{\perp}^2}\delta\! J_{z}=- i\frac{4\pi}{ck_{\perp}}\delta\! J_{\parallel}\;,\\
&& \delta\! B_{z}=0\;,\\
&& \delta\! E_{x}=- i\frac{4 \pi}{\omega}\delta\! J_{x}=- i\frac{4 \pi}{ck}\delta\! J_{\parallel}\;,\\
&& \delta\! E_{y}=0\;,\\
&& \delta\! E_{z}=-\cos{\theta}\delta\!B_{y}=- i\frac{4 \pi
  k_{\parallel}}{ck k_{\perp}}\delta\! J_{\parallel}\;.
\end{eqnarray}\end{mathletters}
The component of the current along the magnetic field in the proper
frame is defined as \begin{eqnarray}
&& J_{\parallel}=\mathbf{J}\cdot\frac{\mathbf{B}}{B}\;.
\end{eqnarray}
The perturbed current can be found from (3.22), (3.24) and (3.14),
\begin{mathletters}\begin{eqnarray}
&& \delta\! J_{x}= i\frac{ck_{\perp}k_{\parallel}}{4 \pi k}\delta\! B_{y}=\frac{k_{\parallel}}{k}J_{\parallel}\;,\\
&& \delta\! J_{y}=0\;,\\
&& \delta\! J_{z}= i\frac{ck_{\perp}^2}{4 \pi k}\delta\!
B_{y}=\frac{k_{\perp}}{k}J_{\parallel}\;,\\
&&\delta\mathbf{J_{\parallel}}= \delta\mathbf{J}= i\frac{c k_{\perp}}{4\pi}\delta\! B_{y}\frac{\mathbf{B}}{B}\;.
\end{eqnarray}\end{mathletters}
From Gauss' law, (3.7), and (3.24) the perturbed charge density is
\begin{eqnarray}
&& \delta\!\rho_{e}= i\frac{k_{\perp}}{4 \pi}\delta\! B_{y}=\frac{J_{\parallel}}{c}\;.
\end{eqnarray}
The second equality in (3.27) utilized the perturbed law of current
conservation which is obtained from the combination of (3.7) and (3.4). The
equations (3.24)-(3.26) can be combined to give the following useful relations amongst dynamical quantities that are propagated along the Alfven characteristics:
\begin{mathletters}\begin{eqnarray}
&& \delta\mathbf{J}\times\mathbf{B}=0\;,\\
&& \delta\mathbf{E}\cdot\mathbf{B}=0\;,\\
&& \delta\mathbf{E}\cdot\delta\mathbf{B}=0\;,\\
&& \delta\mathbf{B}\cdot\mathbf{B}=0\;.
\end{eqnarray}\end{mathletters}Covariantly speaking, (3.28) implies the
coordinate independent relations:\begin{mathletters}\begin{eqnarray}
&& \delta\!J^{\mu}F_{\mu\nu}=0\;,\\
&&\ast\delta\! F^{\mu\nu} F_{\mu\nu}=0\;,\\
&& \ast\delta\!F^{\mu\nu}\delta F_{\mu\nu}=0\;,\\
&& \delta\! F^{\mu\nu}F_{\mu\nu}=0\;,\\
&&\delta F^{\mu\nu}\delta F_{\mu\nu}=0\;,\\
&& \delta\! F_{\mu\nu}U^{\nu}_{_{G}}=0\;,\\
&& F_{\mu\nu}U_{_{G}}^{\nu}=0\;.
\end{eqnarray}\end{mathletters}Equations (3.29a-d) are the coordinate
independent analogs of the local expressions (3.28a-d). From (3.29a),
the current flows along the magnetic field lines in a proper frame
(this expression incorporates the assumption, (3.2), that field line curvature
radii are large compared to the plasma mode wavelength). One can also
get the orientation of the propagation vector relative to the
perturbed and unperturbed fields in all frames by expressing (3.29a)
in terms of the perturbed covariant version of Ampere's Law,
\begin{eqnarray}
&& \delta\! F^{\mu\alpha}k_{\alpha}F_{\mu\nu}=0\;.
\end{eqnarray}
\par \textbf{The Alfven mode transports a charge and a field aligned current (covariantly speaking, a force free current)}.
\section{Causality in Force Free Magnetospheres}The calculation in the
last section allows one to compute the characteristics of short
wavelength waves in a force free magnetosphere. There are four modes:
both ingoing and outgoing fast and Alfven waves. The characteristics
by themselves are not enough to time evolve a force free system, one
must also know what information can be propagated along these
curves. The force free waves carry only two independent pieces of
information. In order to see this, note that the system has ten
parameters: three magnetic field components, three electric field
components, three current components and the charge. There are eight
independent relations connecting these quantities - the force free
equation (2.1) has three constraints, the second order Maxwell's
equation, (3.10), has three constraints as well and there are the two
divergence relations involving the field components in (2.4) and
(2.5). This leaves two independent quantities that are readily
identified from the results of section 3 as the parallel (force free)
current, $\mathbf{J_{\parallel}}$ and the displacement current
orthogonal to the $\mathbf{k}-\mathbf{B}$ plane, $\delta\! E_{y}$.\
\subsection{Fast Wave Characteristics}The fast wave characteristics
can be used to propagate finite changes in the displacement current
orthogonal to the $\mathbf{k}-\mathbf{B}$ plane. Maxwell's equations
can be used to find the corresponding induced magnetic field. By
contrast, no information on the force-free current can be propagated
along the fast characteristics. If one were to cover a region of
spacetime with a mesh of short wavelength fast wave characteristics
(i.e., the boundary of the spacetime region is a essentially a pure
fast wave emitter and absorber), and time evolve the system using the method of characteristics then
one would find that the current and charge in the spacetime would
remain unchanged for all time, in all frames.
\par  There is a claim in \citet{bla01,bla02} that since the fast
wave can transport a toroidal magnetic perturbation, ``it carries
information about the poloidal current.'' This conjecture is
inconsistent with the nature of the fast wave. The fast wave, 
like the light wave carries only displacement current. The time
dependent version of Ampere's law does not require a coupling between physical current and the magnetic field.
\subsection{Alfven Wave Characteristics} By contrast,  changes in the
parallel current can be propagated along the Alfven wave
characteristics. However, no changes in in the displacement current
orthogonal to the $\mathbf{k}-\mathbf{B}$ plane propagate along these
characteristics. From equation (3.27) this current is essentially a
pure charge moving along the magnetic field lines at the speed of
light in the proper frame. The Alfven wave can thought of as a charge
discontinuity or perturbation propagating at the speed
of light along the magnetic field (and its associated parallel
current) and the resulting changes in the electromagnetic field can be
found from Maxwell's equations.
\par Note that in the degenerate case of wave propagation parallel to
the magnetic field, there is no $\mathbf{k}-\mathbf{B}$ plane. In the
degenerate case, (3.24) implies that the Alfven wave carries no
physical current, only a displacement current, $\delta E_{z}$.
\subsubsection{The Force Free Plasma-Filled Waveguide} A very important application of the nature of the Alfven wave
characteristics is a semi-infinite cylindrical waveguide threaded by a
uniform axial magnetic field and filled with a force free plasma. A
rotating conductive disk is suddenly attached to one end of the
plasma-filled waveguide (see Chapter 2 of \citet{pun01} for pictures and
complete details). The conductive disk behaves like a unipolar
inductor or Faraday wheel. When the Faraday wheel is connected, a
Poynting flux begins to be radiated down the plasma-filled
waveguide. The Poynting flux is supported by a force free current. In
the language of force free waves, the Faraday wheel has launched an
axisymmetric force free Alfven step wave (see the Appendix B for details). 
\par Note that by (3.26d), the Alfven wave needs to be oblique if it
supports a field aligned current (i.e., a component of the propagation
vector that is orthogonal to the field, $k_{\perp}$, is required). In
the plasma-filled waveguide, the continuity of the tangential electric
field at the surface of the rotating disk and the degeneracy condition
in the force free plasma imply that there is a poloidal electric
field, $E_{\perp}$, that is launched by the Faraday wheel orthogonal to the poloidal magnetic field, $B^{P}$,
\begin{eqnarray}
&& E_{\perp}=-\frac{\Omega_{_{D}}B^{P}\rho}{c}\;.
\end{eqnarray}The coordinate, $\rho$, is the radial coordinate in a
cylindrical coordinate system and the field line angular velocity,
$\Omega_{_{F}}$ is identical to the disk angular velocity,
$\Omega_{_{D}}$. The radial variation of the electric field in
(4.1) and in the oscillatory modes of (B1) is tantamount to a standing
wave component of the propagation vector that is orthogonal to the
magnetic field (note that the same wave polarization phenomenon occurs in vacuum electromagnetic waveguides). The step
Alfven wave in the cylinder (constructed in Appendix B) is a
superposition (Fourier integral) of the oblique, axisymmetric Alfven
normal modes of the waveguide. The standing wave properties
(obliquity) of the axisymmetric normal modes in the cylindrical
waveguide are a consequence of the boundary conditions on the cylinder walls. 
\subsubsection{Pulsar Magnetospheres} One would expect a strong analogy between the plasma-filled
waveguide and an idealized, force free, field aligned pulsar magnetosphere. The
neutron star is the unipolar inductor, the magnetosphere is the
plasma-filled waveguide and the wind is the Poynting flux in this
analogy. Thus, one would expect that the pulsar wind is
essentially an Alfven wave radiated by the neutron star in this
idealized limit.
\par Unfortunately, the causal nature of MHD in a relativistic
magnetosphere was obfuscated by anecdotal comments in \citet{bla01}
through an analysis in the force free approximation. If the force free
analysis is not pathological then it must agree with the magnetically
dominated limit of MHD. There is a claim that for axisymmetric modes,
the Alfven wave can only support variations in the poloidal magnetic
field, not toroidal variations. However, this cannot be correct as the
example of the plasma-filled waveguide shows by the explicit
construction of the axisymmetric Alfven modes that transport toroidal
magnetic fields. The concept that a standing wave component of the
propagation vector that is orthogonal to the magnetic field can exist
in an axisymmetric mode (as in the plasma-filled waveguide) was not
understood in \citet{bla01}. The existence of such a wave vector
polarization implies, with the aid of (3.30) that
axisymmetric Alfven modes that are capable of transporting toroidal
field perturbations along the direction of poloidal magnetic field
most certainly exist in a pulsar magnetosphere. This oversight allows
for the claim that the force free analysis leads to an ``opposite
conclusion'' to that of the MHD analysis. In reality, if one is
prudent, it can be shown as it has be done here that the two analyses agree, as they must. 
 
\section{Applications to Black Hole Magnetospheres}The force free
approximation has been used in models of black hole magnetospheres in
which spin energy is extracted electromagnetically from the black hole
\citep{blz77,thp86}. The force free wind systems in these treatments
are characterized by two constants in each magnetic flux tube: the
poloidal current (equivalently the toroidal magnetic field or angular
momentum flux) and the field line angular velocity (this is quantity
is equivalent to specifying the ratio of energy flux to angular momentum flux, the Goldreich-Julian charge density and the cross field potential). 
\par The mathematical method of solution implements the force free condition
at the horizon as a boundary condition that is used to determine the
field aligned current density and the field line angular velocity in
each flux tube. Since, by (3.17), the fast wave can propagate in any
direction, at the speed of light, outgoing fast waves can propagate
from anywhere outside of the event horizon (the fast critical surface
for outgoing waves is the event horizon). By contrast, the "light
cylinder" located within the ergosphere is the Alfven critical
surface, which no outgoing Alfven wave can traverse \citep{bla02,pun01}. 
For rapidly rotating black holes, this surface is sufficiently far
from the event horizon that the constraints imposed by the spacetime
metric near the horizon are irrelevant. Thus, the use of the horizon
boundary condition can only be justified from a causal perspective if
the fast mode is involved in determining the wind constants. However,
by (3.19), changes in the poloidal current or charge can not be
transmitted along the fast wave characteristics in the black hole
magnetosphere. Thus, the spacetime near event horizon can not affect
changes in the wind parameters. The use of the event horizon as a
causal wind boundary is not permitted in the
force free approximation. Consequently, the mathematical method of
solution is flawed. The same causal structure is known to exist in
magnetically dominated perfect MHD (see Chapter 6 of
\citet{pun01}). Going to the force free limit was conjectured in
\citep{bla01,bla02} to obviate any need for concern with this method of
solution that were revealed in the full MHD analysis. This claim is
clearly not supported by the results of this article.
\par The causal structure of the black hole magnetosphere in the force free limit has been shown, in this article to behave formally as the zero mass limit of MHD. As in perfect MHD, the wind parameters must be created outside of (farther out from the black hole than) the ergospheric Alfven critical surface. 
\par A further critique of the MHD analysis of black hole causality was
that the Alfven wave really does not transport changes in current and
charge, only oscillations \citep{bla01}. However, this statement is
not an accurate depiction of the nature of propagating disturbances in
MHD. It is well known that small amplitude MHD step waves obey the
same dispersion relations as oscillatory waves. Alfven step waves and
Alfven wave packets in general obey the same constitutive relations
amongst field and dynamical quantities and propagate along the same
characteristics as the oscillatory waves \citep{kap66}. This also
follows by taking the appropriate Fourier integral to create a step
wave from sinusoidal waves. The Alfven wave packets can transport
finite changes in the charge density and field aligned currents in a
force free magnetosphere. This is shown by explicit construction in
Appendix B.
\section{Discussion}There are three main reasons for not considering black hole magnetospheres that are force free
\textbf{everywhere} to be physically realizable:
\begin{enumerate}
\item As mentioned in the Introduction, the physical plasma state near
  the event horizon is always inertially dominated. This physical state is exactly the opposite of the zero inertia limit that is inherent to the force free
assumption.
\item From equation (2.7) there is no region within the force free
  magnetosphere where energy flux (Poynting flux) and angular momentum flux can be created. Energy and angular momentum flux (essentially, the poloidal current and Goldreich-Julian charge density) must be injected into the wind system at the boundaries. If the magnetosphere is force free everywhere then the energy flux (current and charge) must emerge from the horizon. However, the charge 
and current necessary to support the energy and angular momentum flux
can not be propagated along the plasma wave characteristics emanating
from the spacetime near the horizon as discussed in the last
section. Thus, such a wind system is inherently acausal.
\item There are three MHD computer simulations of magnetically
  dominated black hole magnetospheres (the magnetic energy density
  exceeds the plasma energy density in the initial state) that have
  been published. All of them evolve to a state with strong
  cross-field (inertial, not force free) current flow in the ergosphere \citep{cak00,kom02,sem01}. The inertial current flow
initiates in regions far enough from the event horizon that it is not
interpretable as a horizon surface current. The strong cross-field
current is a necessary component for extracting black hole energy in the theory of black hole gravitohydromagnetics
(GHM) as well in \citet{pun01}.
\end{enumerate}
\par It should be noted that a wind system beyond (farther from the
black hole than) the ergospheric dynamo region for the toroidal magnetic field (poloidal current) could in principle be well described by the force 
free approximation. From the nature of the force free characteristics,
this dynamo region can exist only in the spacetime farther from the
black hole than the ergospheric light cylinder (Alfven critical
surface). The Poynting flux can propagate outward from the dynamo
region to asymptotic infinity as an essentially force free wind. Such
a wind system is described in detail in Chapter 9 of \citet{pun01}.

\acknowledgments
I would like to thank Ferd Coroniti for his valuable and intelligent insights. I am also indebted to Serguei Komissarov for his useful comments regarding force free plasmas and propagating discontiuities.
\appendix
\section{First Order Curvature Effects - Hybrid Modes} In this
appendix, we calculate the first order effects of field line curvature
on the force free small amplitude plasma modes in a proper frame. One
finds new hybrid modes as in MHD. The first order corrections to the
short wavelength approximation can be found in the eikonal approximation. The sinusoidal solutions described in (3.2) are replaced by:\begin{equation}
J^{\mu}=J^{\mu}_{_{0}} +\delta J^{\mu} e^{\imath[\int\!\mathbf{k}\cdot
  d\mathbf{r} - \omega t]}\;,
\end{equation} where $\mathbf{k}$ is a slowly varying vector valued function. 
In the present circumstance, the unperturbed current no longer vanishes by assumption. However, by the force free condition, it flows parallel to the magnetic field in the proper frame. This can be expressed in terms of a slowly varying ``radius of curvature'', $R$, as
\begin{equation} \mathbf{J}_{_{0}} = c R^{-1} \mathbf{B}_{_{0}} \;.
\end{equation}
The validity of the eikonal approximation in (A1) requires that:
\begin {eqnarray}
&\mid \mathbf{k} \mid R\gg 1\;,\\
&\mbox{\Large{$\frac{\mid \mathbf{\nabla} k \mid}{k^{2}}$}}\ll 1\;.
\end{eqnarray}
\par
The perturbed Maxwell's equations and the perturbed degeneracy condition carryover from the sinusoidal case, (3.4) - (3.11), unchanged. The only perturbation equation that changes is the force free condition, (3.2), that now becomes:
\begin{equation}
\mathbf{B}_{_{0}} \times \delta\mathbf{J} + \delta\mathbf{B}\times \mathbf{J}_{_{0}}=0\;.
\end{equation}
The perturbed force free relation found from expanding the "x" and "z" components of (A5)  is \begin {equation}
\frac{4\pi ik^{2}c^{2}}{c^2 k^{2} - \omega^{2}}(kR)^{-1} \delta J_{z} = \delta J_{y}\;.
\end {equation} Note that by (A3), the large radius of curvature approximation, equation (A6) is approximately (3.12). The "y" component of the perturbed force free relation (A5) combined with (A6) yields the relation:
\begin{equation}
\frac{16\pi^{2} k^{4} c^{4}(kR)^{-2} - \left[ c^{2}k^{2} - \omega^{2} \right]^{2}}{\left[ c^{2}k^{2} - \omega^{2} \right]^{2}}\frac{k_{\parallel}}{k} B_{_{0}}\delta J_{z} + \frac{k_{\perp}}{k} B_{_{0}}\delta J_{x} = 0\;.
\end{equation} Note that in the limit of a large magnetic field radius of curvature that (A7) becomes (3.13). Thus, the complete set of perturbed equations reduce to the equations found in section 3 as the curvature of the field gets small.
\par The dispersion relation is found by combining (A7) with the perturbed degeneracy condition (3.14),
\begin{equation}
\left[ \omega^{2} - c^{2}k^{2}\right]\left[ \omega^{2} - c^{2}k_{\parallel}^{2}\right]= 16\pi^{2} k^{2} k_{\parallel}^{2} c^{4} (kR)^{-2}\;.\end{equation} There are two roots of the dispersion relation.:\begin{eqnarray}
&\omega^{2}=c^{2}k^{2} + \frac{1}{2}c^{2}\left[\sqrt{k_{\perp}^{4} + 64\pi^{2} k_{\parallel}^{2}k^{2}(kR)^{-2}} - k_{\perp}^{2}\right]\;,\\
&\omega^{2}= c^{2}k_{\parallel}^{2} - \frac{1}{2}c^{2}\left[\sqrt{k_{\perp}^{4} + 64\pi^{2} k_{\parallel}^{2}k^{2}(kR)^{-2}} - k_{\perp}^{2}\right]\;.
\end{eqnarray} Differentiating (A9) and (A10) yields the group velocities of the two modes. For parallel propagation, the group velocity is the speed of light up to terms that are $O\left[ (kR)^{-2}\right]$. The parallel propagation assumption is invalidated at second order since the magnetic field is curved. In the limit that $(k_{\perp} R)^2\gg 1$, one can obtain the following Taylor series expansions of the group velocity for oblique propagation:
\begin{eqnarray}
&\mbox{\Large{$\frac{d\omega}{dk}$}} \approx\pm\mbox{\Large{$\frac{c}{\sqrt{1+16 \pi^{2}
    \cot^{2}\theta (kR)^{-2}}}$}}\approx \pm c \left[ 1 - 8 \pi^{2}(kR)^{-2} \cot^{2}\theta\right]\;,\\
&\mbox{\Large{$\frac{d\omega}{dk}$}} \approx\pm\mbox{\Large{$\frac{c \cos\theta}{\sqrt{1-16
    \pi^{2} \cot^{2}\theta (kR)^{-2}}}$}}\approx \pm c \cos\theta \left[ 1 + 8 \pi^{2}(kR)^{-2} \csc^{2}\theta\right]\;.
\end{eqnarray} Equations (A9) and (A11) approximate to (3.16) and
    (3.17) as the radius of curvature becomes infinite. Thus, these
    equations represent the fast hybrid mode. Similarly, (A10) and
    (A12) become (3.22) and (3.23) in the short wavelength limit. These
    equations represent the hybrid Alfven wave. Notice that the fast
    hybrid mode propagates slower than the speed of light for oblique
    waves propagating across a curved background magnetic field. From
    (A12), the Alfven mode does not propagate perpendicular to the
    field. Thus, there is not Alfvenic properties for the fast mode to
    couple into this case. Thus, the hybrid fast mode still propagates
    at the speed of light for perpendicular propagation as indicated
    in (A11).
\par There are implications of this result to the primary focus of this article. Since the fast wave no longer propagates isotropically, it can now carry a physical current (not just a displacement current). Denote the displacement current as, $J^{y}_{_{D}}$. Then (3.11a), (A6) and (A9) imply that the ratio of displacement current to physical current for oblique fast hybrid waves is:
\begin{equation}
\frac{J^{y}_{_{D}}}{J_{\parallel}}\approx
\frac{(kR)^{3}\tan^{4}\theta\sin\theta}{64\pi}\;.\end{equation}
Consequently, the ratio of displacement current to the physical
current diverges in the short wavelength limit as in
(3.19). Similarly, from the perturbed Gauss' law (3.7a), (3.11c) and
(A9), one can find the ratio of displacement current to the charge
density:
\begin{equation}
\frac{J^{y}_{_{D}}}{\rho_{e}}\approx c
\frac{(kR)^{3}\tan^{4}\theta\sin\theta}{64\pi}\;.\end{equation}
Comparing (A9) to (A13) and (A14), note that as the fast wave
propagation speed slows down due to field line curvature, it carries
more physical current and charge. It becomes increasingly more like
the Alfven wave as the field line curvature increases. However, (A13)
and (A14) show that in the eikonal approximation, the fast wave is
still almost purely electrodynamic: a propagating displacement current
with a minute physical current that is only $O\left[ (kR)^{-3}\right]$ of the displacement current.
\section{The Alfven Wave Packet in the Cylindrical Waveguide}In this
appendix it is shown by explicit construction that a superposition of
oscillatory Alfven waves (an Alfven wave packet) does indeed transport
field aligned currents and charge, not just oscillatory
perturbations. We rely on the magnetically dominated MHD treatment of
the plasma-filled cylindrical waveguide found in Chapter 2 of \citet{pun01} and
pass to the force-free limit. In order to save space, we do not
reproduce the results of Chapter 2 here, but refer to them in detail and the
reader is urged to consult that reference. In summary, a cylindrical
waveguide filled with a tenuous plasma is placed inside of an
infinitely long solenoid so that it is threaded by a strong axial
magnetic field. The waveguide is semi-infinite in the sense that one
end is terminated by a rotating conductive wheel. This rotating
conductor or Faraday wheel (also known as a homopolar generator or
unipolar inductor) is an active element in the waveguide
circuit that radiates plasma waves. In the first subsection, we
construct the relevant axisymmetric, m=0, perfect MHD Alfven modes in
the cylinder using the detailed results of \citet{pun01}. In the next
subsection, we compare and contrast a wave packet (a step wave) created from a
linear superposition of the previously calculated oscillatory Alfven
modes to the radiation emanating from the Faraday wheel
(the nature of this radiation was derived previously in terms of a formal MHD wind solution
in the waveguide in \citet{pun01}). Finally, it is noted that the
depiction of the wind as a pure Alfven step wave becomes increasingly
more accurate as one passes to the force-free limit.
\subsection{Components of the Oscillatory Waves}This section
constructs an oscillatory solution from Chapter 2 of \citet{pun01} for 
an axisymmetric Alfven wave, $m=0$, in the plasma-filled waveguide
that is consistent with the attachment of the Faraday wheel (in the
realm of magnetically dominated perfect MHD).
\subsubsection{The Electric Field}
\par From equation (2.93b) of \citet{pun01}, the radial electric field
in an axisymmetric Alfven wave in a cylindrical waveguide is arbitrary 
and is determined by the boundaries (the field lines can shear
relative to each other, hence these modes are often called ``shear
Alfven waves''). Considering the frozen-in
condition within the Faraday wheel and the continuity of the tangential 
electric field at the Faraday wheel/plasma interface let us choose the
radial electric field in the Alfven wave to be (in cylindrical coordinates):
\begin{equation}
E^{\rho}=-\frac{\Omega_{_{D}}\rho}{c}B_{_{0}}e^{i(k_{z}z-\omega t)}\;.
\end{equation}
The other electric field components are given in \citet{pun01} by (2.86c)
and (2.93a) as
\begin{eqnarray}
&& E^{\phi}=0\;,\\
&& E^{z}=0\;.
\end{eqnarray}
\subsubsection{The Currents}
From equations (B1)-(B3) and Gauss' law, the charge density is;
\begin{equation}
\rho_{e}=-\frac{\Omega_{_{D}}B_{_{0}}}{2\pi c}e^{i(k_{z}z-\omega t)}\;.
\end{equation}
From the Alfven wave dispersion relation, (2.82) of \citet{pun01},
and the frozen-in form of the momentum equation, (2.86a) of
\citet{pun01}, along with equation(A1), the cross-field current density is
\begin{equation}
J^{\rho} =i\frac{v_{_{I}}k_{z}\Omega_{_{D}}\rho}{4\pi U_{_{A}}^{2}c}B_{_{0}}e^{i(k_{z}z-\omega t)}\;,
\end{equation} where $v_{_{I}}$ is the intermediate three speed and
$U_{_{A}}$ is the pure Alfven speed. In the force free limit, the pure
Alfven speed diverges as the intermediate speed approaches the speed
of light.
From the law of current conservation,
\begin{equation}
\frac{\partial\rho_{e}}{\partial t}+\mathbf{\nabla}\cdot\mathbf{J}=0\;,
\end{equation}combined with (B4) and (B5) yields the axial current density:
\begin{equation}
J^{z}=\frac{\rho_{e}}{v_{_{I}}}c^{2}=-\frac{\Omega_{_{D}}B_{_{0}}c}{2\pi v_{_{I}}}e^{i(k_{z}z-\omega t)}\;.
\end{equation}
\subsubsection{The Magnetic Field}
Using the value of the axial current in (B7) in Ampere's law yields the toroidal magnetic field strength:
\begin{equation}
B^{\phi}=-\frac{B_{_{0}}\Omega_{_{D}}\rho}{v_{_{I}}}e^{i(k_{z}z-\omega
  t)}\;.
\end{equation}The other components are given by (2.92a) and (2.96) of \citet{pun01}:
\begin{eqnarray}
&& B^{z}=0\;,\\
&& B^{\rho}=0\;.
\end{eqnarray}
\subsection{Making Wave Packets}We can construct wave packets of the
oscillatory solutions above by taking Fourier integrals. We are interested in the step wave. For the Alfven wave, one can use the dispersion relation
(2.82) of \citet{pun01} in the cylinder to write,
\begin{equation}
e^{i(k_{z}z-\omega t)}=e^{-ik_{z}(v_{_{I}}t-z)}\;.
\end{equation}The step function is given by,
\begin{equation}
\Theta(v_{_{I}}t-z)=-\frac{1}{2\pi i}\int_{-\infty}^{+\infty}\frac{e^{-ik_{z}(v_{_{I}}t-z)}dk_{z}}{k_{z}+i\epsilon}\;.
\end{equation}We construct a wave packet of Alfven waves,
$\Psi$, with a spectral amplitude, $A(k_{z})$, 
\begin{eqnarray}
&& \Psi=\int_{-\infty}^{+\infty}A(k_{z})\psi(k_{z})dk_{z}\;,\\
&& A(k_{z})=-\frac{1}{(2\pi i)(k_{z}+i\epsilon)}\;,
\end{eqnarray}where $\psi(k_{z})$ is one of the oscillatory
wave function components in equations (B1)-(B10) and $\epsilon$ is an
arbitrarily small positive number in the usual sense. The wave packet
of oscillatory solutions with field and current components given by
(B1)-(B10) and spectral amplitude given by (B14) has the following
field and current distributions (for the sake of demonstrating
equivalence, the corresponding equation number from \citet{pun01} that
were derived by the MHD wind calculation in the cylinder is placed on the right hand side of the equals sign in these relations):

\begin{eqnarray}
&& E^{\rho}=-\frac{\Omega_{_{D}}\rho}{c}B_{_{0}}\Theta(v_{_{I}}t-z)=(2.98)\;,\\
&& E^{\phi}=0\;,\\
&& E^{z}=0\;,\\
&& \rho_{e}=-\frac{\Omega_{_{D}}B_{_{0}}}{2\pi c}\Theta(v_{_{I}}t-z)=(2.99)\;,\\
&& J^{\rho} =\frac{v_{_{I}}\Omega_{_{D}}\rho}{4\pi U_{_{A}}^{2}c}B_{_{0}}\delta(v_{_{I}}t-z)\;,\\
&& J^{z}=\frac{\rho_{e}}{v_{_{I}}}c^{2}=-\frac{\Omega_{_{D}}B_{_{0}}c}{2\pi v_{_{I}}}\Theta(v_{_{I}}t-z)=(2.119)\;,\\
&& B^{\phi}=-\frac{B_{_{0}}\Omega_{_{D}}\rho}{v_{_{I}}}\Theta(v_{_{I}}t-z)=(2.118)\;,\\
&& B^{z}=0\;,\\
&& B^{\rho}=0\;.
\end{eqnarray}These quantities agree with the downstream state of the
plasma that was found near the flow front in \citet{pun01} in the limit that $v_{z}=v_{_{I}}$. Notice that $J^{\rho}$ vanishes downstream of the wavefront and
only has a surface component on the wavefront as it does in equation
(2.110) of \citet{pun01}.
\subsection{Conclusion} From equations (B18) and (B20), the Alfven wave
packet actually transports charge and field aligned current. 
\par The solution is not exact. There are errors associated with the
small inertial terms in the magnetically dominated limit. At the
wavefront, the relativistic MHD shock equations do not solve
exactly. These equations are the conservation of the axial components
of the stress energy tensor across the discontinuity in the frame of
the wavefront. Pressure balance (magnetic pressure from the toroidal
magnetic field in the frame of the wavefront) can not be achieved
because the Alfven wave, unlike the fast wave, has no compressive
properties. The errors in the stress-energy balance can be found in
the frame of the propagating discontinuity by using (B15) and (B21) to
compute the proper toroidal magnetic field. The errors in the shock
equations are on the order of $(\Omega_{_{D}}\rho
U_{_{A}}^{-1})^2$. This is biquadratic in two small quantities: $\Omega_{_{D}}\rho / c \ll 1$ by construction and by
the magnetically dominated condition, $c U_{_{A}}^{-1} \ll 1$. The
violation of the shock relations are therefore a very small second
order effect in the magnetically dominated limit. Notice that the
errors vanish completely in the force-free limit as the Alfven three
speed approaches the speed of light. In magnetically dominated perfect
MHD, these small errors are accounted for by a fast switch-on shock
that creates a $B^{\rho}$ and an $E^{\phi}$ downstream of the shock
front. This is an MHD precursor, an infinitesimal distance upstream,
to the Alfven rotational discontinuity at the terminus of the Alfven
step wave. As demonstrated by explicit construction above, it is the
Alfven rotational discontinuity that imprints the charge and field
aligned current on the waveguide plasma. The interpretation of the
waveguide wind solution as an Alfven wave is exact in the force-free
limit and is extremely accurate to first order in the magnetically dominated MHD limit. However, in relativistic MHD it is important to note that both fast and Alfven modes are needed to solve this type of ``Riemann problem.'' 
\par It was an oversight in \citet{pun01} that nonlinear fast waves
such as an abrupt discontinuity can in principle transport charge and
current. However, in practice, due to the limited polarization
properties of such waves (the fast discontinuity can only affect
changes in the magnetic field in the plane spanned by the normal to
the wavefront and the magnetic field upstream) unless the  magnetic field is
perfectly homogenous (not the case in a magnetosphere) they must be
accompanied by an Alfven rotational discontinuity in a magnetosphere
in order for the discontinuities to be solutions of the relativistic
shock equations \citep{kap66}. The important point is that in general
both modes are needed. The discussions of \citet{pun01} were designed
to show that the Alfven wave is always required in MHD, so that the fast wave
cannot determine the wind constants alone - and this conclusion is
unchanged and hence the MHD causality arguments are unchanged. In the
force-free limit, the situation is more extreme. Since the fast wave
is linear in the force free limit (the fast speed is always c), all
wave packets constructed from the complete set of oscillatory fast
modes will also carry no current or charge in any frame (see section
3.1). Thus, the Alfven wave is responsible for setting the wind
constants in a force free black hole magnetosphere.
\par 
This discussion corroborates the physical accuracy of the assumptions
used in the one dimensional string analysis of flux tubes in
\citet{sem01}. They actually ignore the fast mode completely (because
as they say, the fast wave is well known to primarily provide only
pressure changes in an MHD flux tube) and only consider the Alfven
mode. This seems like a very reasonable assumption based on the
analysis provided here.

\end{document}